\documentclass{iau}
\usepackage{graphicx}

\title[TBN]
{Interferometry meets the third and \\ fourth dimensions in galaxies}

\author[Virginia Trimble]
{Virginia Trimble}

\affiliation{University of California Irvine, \\ 
Department of Physics and Astronomy \\ Irvine, California 92697, USA \\ email: {\tt vtrimble$@$uci.edu} \\[\affilskip]}

\pubyear{2014}
\volume{} 
\pagerange{\? - \?}
\jname{TBN  \\  TBN}
\editors{A.C. Editor, B.D. Editor \& C.E. Editor, eds.}
\begin{document}

\maketitle

\begin{abstract}
Radio astronomy began with one array (Jansky's) and one paraboloid of revolution (Reber's) as collecting areas and has now reached the point where a large number of facilities are arrays of paraboloids, each of which would have looked enormous to Reber in 1932.  In the process, interferometry has contributed to the counting of radio sources, establishing superluminal velocities in AGN jets, mapping of sources from the bipolar cow shape on up to full grey-scale and colored images, determining spectral energy distributions requiring non-thermal emission processes, and much else.  The process has not been free of competition and controversy, at least partly because it is just a little difficult to understand how earth-rotation, aperture-synthesis interferometry works.  Some very important results, for instance the mapping of HI in the Milky Way to reveal spiral arms, warping, and flaring, actually came from single moderate-sized paraboloids.  The entry of China into the radio astronomy community has given large (40-110 meter) paraboloids a new lease on life.
\keywords{Interferometry, Radio Astronomy, Superluminal Velocities, VLA, ALMA}
\end{abstract}

\section{Introduction}
Critics have claimed that astronomy is not a science, because you cannot repeat an experiment changing one variable at a time.  At least part of the answer is, no, but we can find stars that differ from our sun only in mass, or age, or metal content, or rotation rate, or magnetic field and carry out comparisons that achieve the same goal.  This works for stars because most of them are nearly spherical.  Manifestly that is not true for galaxies of most types.  Thus the possibility of determining three-dimensional structures for galaxies and correcting for orientation addresses the question of whether galactic astronomy can claim to be science.

A seemingly-simple example is to extract the real distribution of elliptical shapes from the statistics of eccentricities on the plane of the sky, assuming that a large sample will be randomly oriented relative to us.  That assumption is clearly not true in a magnitude-limited sample, because an end-on prolate spheroid will look brighter than a face-on oblate spheroid covering the same area of the sky.  Triaxiaity, existence of thick disk galaxies, bars, and dominance of dark matter (whose shape is not necessarily traced by luminous matter any more than its density is) all make this worse.

Three dimensional studies of galaxies and their parts generally address much more detailed questions than ``Is galactic astronomy science?''â and correspondingly have much better chances of answering the questions asked.  The role of interferometry has generally been to enhance angular and occasionally spectral resolution, not without other costs, and has generally meant radio and millimeter interferometry when applied on galactic or Galactic scales, though one early design for Gaia, called GAIA-ROEMER was intended to have interferometric capability.

The oral version of this presentation included images of a very large number of radio telescopes, some interferometers, some monoliths, and some not quite sure.  The cited references not specifically mentioned in the text for a particular idea or result are the ones from which most of those images were taken.  Krauss (1986) was a particularly rich source.  Kellermann \& Moran (2001) have provided a truly expert overview of the evolution of radio interferometry from Ryle \& Vonberg's (1946) 0.5 km, two-element Michelson-analog device up to the epoch of publication and indeed somewhat beyond, with discussions of larger arrays, space interferometry, and moving further toward short wavelengths in the future.  They do not mention that Thomas Gold (Gold \& Mitton 2012, p. 102) believed that he was the first in 1955 to suggest using separate, very accurate clocks at widely separated antennas to permit baselines as large as the earth.

\section{What is an interferometer?}
The primordial version was the 1801 Young two-slit experiment, which demonstrated the wave nature of light, and, as in his case, if you see fringes, it means your interferometer is working.  Fizeau thought of applying interferometric concepts in astronomy in 1868, and Stephan at Marseilles in 1874 concluded that stars have angular diameters less than 0.158 arc sec.  He masked the middle of a mirror, as did Michelson at Mt. Wilson to measure Jupiter's moons in 1891.  Michelson extended the diameter of the Mt. Wilson 100$^{\prime \prime}$  and Pease \& Anderson used this kluge to find a diameter of 0.047 arcsec for Betelgeuse in 1920.  All of these are, in effect, time reversals of the Young experiment.  Optical interferometry is undoubtedly a growth industry, but has, so far, been applied largely to stars rather than galaxies (Lindemann 2011).

Fabry-Perot interferometers are the exception.  In these, two flat glass (etc.) surfaces very close together bounce light back and forth and so pick out some one particular wavelength with constructive interference.  They can be tuned by changing the spacing. Thus the first successful application of interferometry in astronomy came in 1910 when Fabry \& Buisson (1914) examined the Orion Nebula with a small telescope at Marseilles (Lequeux 2013).  Their goal was to get a sufficiently precise wavelength for an emission feature near $5007 \AA$ to permit its identification with some laboratory substance.  In this they failed and came to the end of their paper still calling it Nebulium.  Emission lines at $3726.1$ and $3728.8 \AA$ also remained unidentified.  The line widths (a real achievement of the F-P method) suggested atomic weights between those of hydrogen and helium, and they suggested identification with two such elements that had been postulated by Rydberg.

It is not true that F-P applications belong exclusively to Marseilles (my thesis advisor was very fond of them), but Marcelin et al. (1983) reported a very interesting scan through the width of the $H \alpha$ feature in the nearly face-on spiral NGC 2903.  Different gas blobs show up at different velocities, and we are presumably seeing some combination of inflow, outflow, and galactic fountain processes.

Most of the rest of galactic interferometry has occurred at radio wavelengths, increasingly with the use of aperture synthesis and gradually pushing to shorter wavelengths.

\section{What is a dimension?}
Figure 1 is a perfectly possible two-dimensional image, but you cannot instruct one in 3-d.  It is called a three-pronged blivet, and in one-d it would be merely a line of finite width.  The standard one-dimensional joke is, however, the story of the engineer who was called into consultation at a dairy farm and began by saying ``Consider a spherical cow.''  This works for stars (where 1-d is the radius), and T, P, $\rho$, and composition can be written as functions of $r$ alone.  But even a spherical-looking galaxy almost certainly has $\sigma(v)$ - the equivalent of temperature - that is unlikely to be isotropic, so 1-d will not do.

\begin{figure}
\begin{center}
\includegraphics{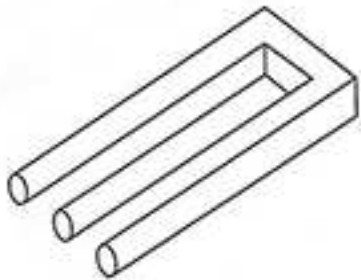}
\caption{A three-pronged blivet.  The radio interferometric equivalent is a map that turns out with negative flux in some places.}
\label{fig1}
\end{center}
\end{figure}

Two dimensions could be either $r$ and $\theta$ or $r$ and $z$ and provide a plausible next step for study of rotating and/or magnetic stars (where angular momentum is apparently conserved on cylinders during evolution).  This still won't do for galaxies, where light clearly does not trace matter, and Schwarzchild's (1982) scheme was therefore doomed to failure.  His idea was this:  assume a potential; put down a large number of point stars and let them orbit in that potential (in circular, radial, banana, and many other orbit shapes); spread the mass of each star around its orbit, proportionate to the dwell time at each point; now add up the masses and see whether they reproduce the potential.  They won't.

The bipolar cow is a bit more promising.  Very many astronomical images do look somewhat like what you would expect if a rapidly rotating Maclauren spheroid distorted unstably into unequal lobes and then split.  I showed 10 images of this general sort, interferometric ones of both interacting galaxies and star formation regions and a couple of simulations.  Fred Adam's drawing of the bipolar cow appears as Fig. 1 of Trimble (2002) and also has the morphology of two blobs with some tenuous connection.  Again as in the astronomical case, higher angular resolution reveals more features, whether disks, arms, jets, and gas streams or ears, hooves, and tails.

\section{Earliest radio astronomy and the Milky Way}
Jansky's (1932) device was a sort of array, and Reber (1940), the world's first dedicated radio astronomer, initially built a paraboloid of revolution, though many of his later devices were arrays and interferometers, often at very low frequency.  Both pioneers observed continuous emission and ended up with maps dominated by the Milky Way.  If you already knew that there is a substructure, a warp, and outer flaring, you can probably see them in their maps.  The next continuum steps, separating out disk and corona and recognizing the North Polar Spur were already dominated by interferometer results from Cambridge and Australia (Pawsey 1964).  And some bright spots were arms end-on. 
 
In contrast, the 21 cm maps of neutral hydrogen emission that firmly established our existence as a spiral galaxy came from paraboloids in Dwingeloo and Sydney (Kerr \& Westerhout 1964).  This was also true for the evidence for disk warping (up to north of sun-center line, down to the south) and flaring at galactic radii larger than the solar circle.  Many (not all) other spiral/disk galaxies have similar features, and very many of the best maps have come from the VLA.  Another interferometer, BIMA (Berkeley-Illinois-Maryland array, now part of CARMA) provided HI velocity maps confirming a Galactic central bar, whose existence had been found first from star counts.

\section{My paraboloid can lick your interferometer (or perhaps conversely)}
Radio astronomy grew out of World War II radar technology and was nurtured by people who were not typically astronomers and who had different vocabulary, goals, and mathematical and engineering skills.  The world was never quite so firmly bifurcated between single large dishes and arrays as the N-S split in German-speaking Europe between Gute Morgen and Gr\H{u}ss Gott or as impenetrable as the E-W split between the parts of eastern Europe where only Pepsi Cola was available and the land of the Coke.  But even the US, which came late to the game, began with a single 1953 dish at Harvard and a pair of 90-foot dishes at Ovens Valley, belonging to Caltech, who had had the excellent sense to import John Bolton from Australia.

Important starting points in the Netherlands were the use of radio-telephone to communicate with distant colonies in the 1930Õsâ and W\H{u}rzburg radar antennas left behind at the end of the war (Strom 2005).  In contrast, Australia had already been using Lloyd telescopes (or sea-cliff interferometers) for radar monitoring during the war. The two beams that interfere are the one coming directly to your dish (or Yagis or whatever) from the source and one reflected off ocean water a few hundred feet below the cliff where your telescope sits.  A big part of the fun of reading about the beginnings of Australian radio astronomy is the wonderful set of names of their field stations:  Badgerys Creek, Dapto, Fleurs, Potts Hill, Dover Heights, Hornsby Valley, and more.  (Orchiston \& Slee 2005).

Fairly clearly, the first successful use of radio interferometry for astronomical purposes happened in Australia in 1945-46.  The target was the sun, and some association of radio emission with active surface regions was quickly established.  Joe Pawsey was the senior member of the group (at age 37), but a great deal of the work was done by Ruby Payne Scott, though she never made the transition to galactic and extragalactic radio astronomy (Goss \& McGee 2010).

By the early 1950s, both the Australian stations and UK ones (that is, Cambridge and Jodrell bank) had an assortment of facilities engaged in providing accurate positions (for optical identifications), source fluxes (for counting mostly), the first, poorly resolved maps, and spectra (for initial studies of emission mechanisms).  Mainstream astronomy was, perhaps, rather slow to accept radio astronomy in general (Jarrell 2005 and many other sources), but interferometry presented some special difficulties.  Figures 2 and 3 illustrate these in frivolous and serious ways.  Figure 2 embodies a remark by Peter A.G Scheuer (the first theorist attached to the Cambridge group) that ``interferometry is like being led blindfolded up a single path to the top of the mountain, and then being asked to describe the entire mountain and its Fourier transform.''  Figure 3 (adapted from Krauss 1986) shows both algebraically and geometrically the relationship between brightness temperature on the sky (in right ascension and declination for instance) and where your dishes are on the ground, the u-v plane.  For a fixed baseline, rotation of the earth will trace out for you some sort of arc in the u-v plane, and the more thorough the uv coverage, the more you will learn about $T_{b} (\alpha, \delta)$.  N antennas will give you $N(N-1)/2$ arcs in the uv plane.  And the enormous achievements of the VLA, other similar arrays, and now ALMA arise from careful choice of antenna spacings and the ability to vary those spacing, until the VLA looking at Sgr B2 and ALMA looking almost anywhere in its field of view fill the uv plane like squashed spiders (Wilson 20133, pp 255 \& 268).

Early radio maps nearly always showed $T_{b}$ contour lines, unfamiliar to optical astronomers, and the transition to grey-scale (or even colored) images that has accompanied the filling up of the UV plane has made those maps look very much more like astronomical images at other wavelengths.  This is an aspect of the history that does not seem to have been much mentioned.

In contrast, the propensity of interferometers to resolve out diffuse flux and to ``confuse'' two or more adjacent weak sources into a single, brighter one, have been extensively discussed (Kellermann \& Moran 2001).  The astronomical context in which confusion in this technical sense caused much anger was the counting of radio sources as a function of the flux received.  Because volume scales as distance cubed and the flux you receive will scale as (distance)$^{-2}$, a homogeneous, isotropic, non-evolving source population will give you a number of source, N, seen down to flux, S, varying as $N(S) \propto S^{-\frac{3}{2}}$.  Steeper relationships mean more sources in the past (for an expanding universe) and flatter ones fewer sources in the past.  There is no doubt that the early Cambridge reports of $N(S)$ were steeper than reality, and the sorting out took several years, with input from Australia, Jodrell Bank, and elsewhere, and with hard feelings left many places (Gold \& Mitton 2012, Longair 2006).

\begin{figure}
\begin{center}
\includegraphics{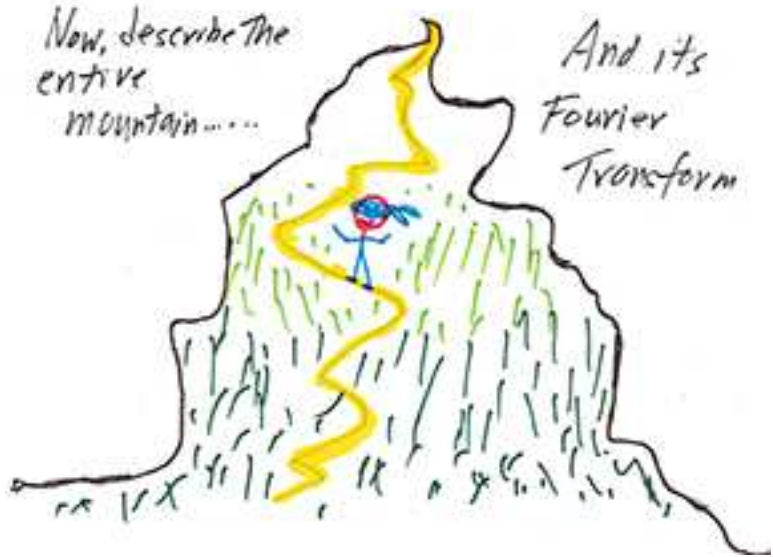} 
\caption{P.A.G. Scheuer's informal 1968 definition of two-element, earth-rotation, aperture-synthesis interferometer.  (author's drawing)}
\label{fig2}
\end{center}
\end{figure}

\begin{figure}
\begin{center}
\includegraphics{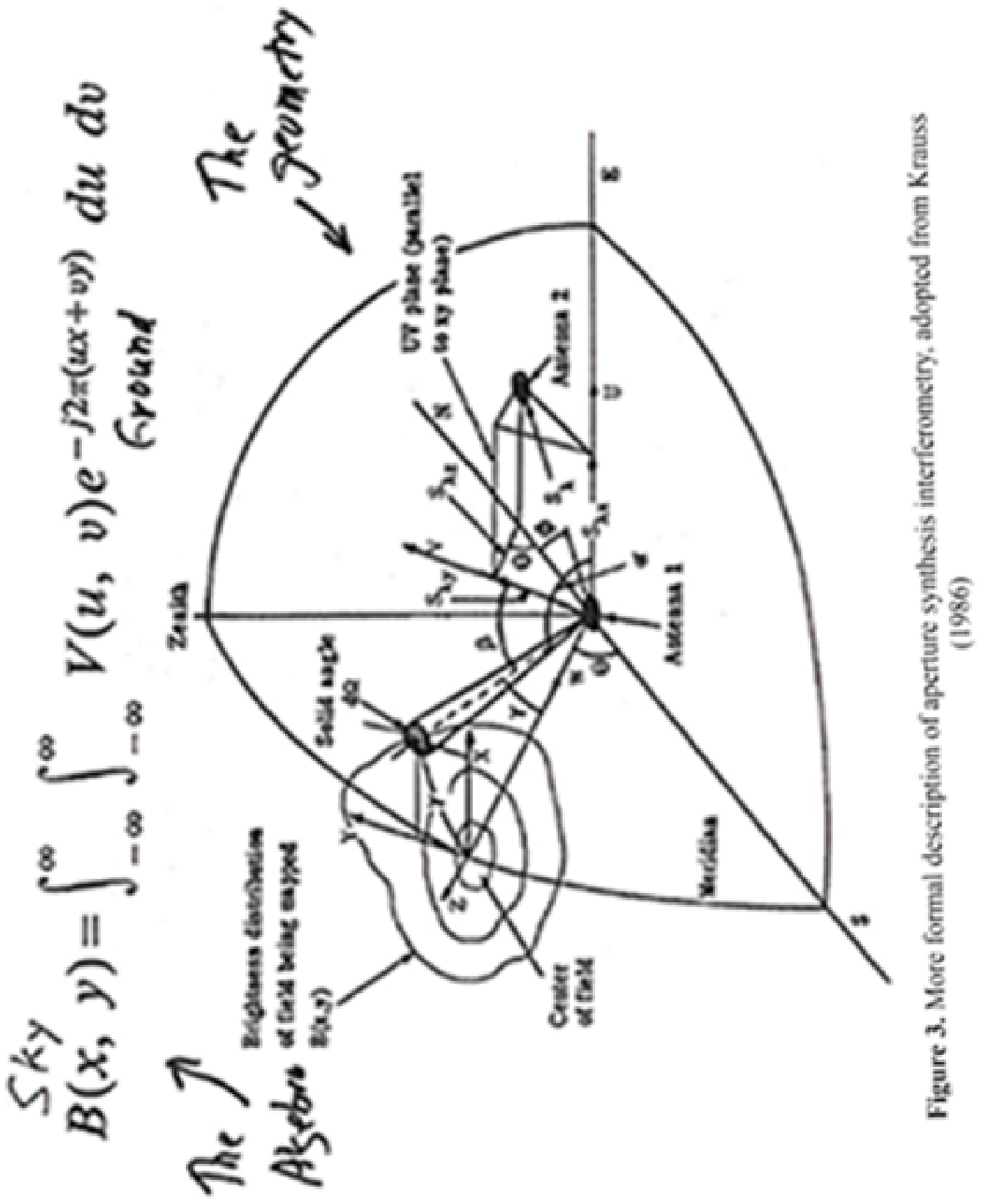} 
\caption{}
\label{}
\end{center}
\end{figure}

There are, to this day, single-paraboloid people and interferometer people, though I think the latter are winning.  The Dwingeloo 25-m is being reconditioned, but the Parkes 64-m (as part of the ATNF), the Nobeyama 45-m, the Lovell 76-m, the Effelsberg 100m, the Greekbank 140-foot, the Algonquin 46-m, and others continue to soldier on.  A few years ago, I would have bet that the 100-meter Greenbank Telescope (the Byrd in the hand), completed in 2000 and the largest fully-steerable dish at present, would be the last of its breed.  But in the last few years, China has commissioned 40, 50, and 65-meter dishes with 110-meter fully-steerable planned.  FAST in Guizhou, their successor to Arecibo (that is a hole in the ground with radio-reflecting surfaces and receivers), will exceed the 1000-foot diameter of the Puerto Rico telescope.

As has happened many times before in radio astronomy, all bets are off when a new player enters the game!

\section{One very important VLBI discovery and the havoc it wrought}
Very Long Baseline Interferometry generally means the sort where the several collecting areas (from 2 to 13 or more) cannot be connected by cables or even radio links, and the first versions used only two or three antennas.  Now the correct explanation for the enormous powers emitted by quasars, radio galaxies, and other active galactic nuclei had been out there since 1964:  accretion onto a very massive, compact central object, according to papers by E.E. Salpeter and by Ya. B. Zel\'dovich and I.D. Novikov.  But a VBI result confused the story for several years.

The quasar 3C279 ($z=0.538$ and distance debated by hold-out supporters of steady state cosmology and non-cosmoloical redshifts) and radio galaxy 3C 120 ($z = 0.033$ and distance not doubted) both had been displaying rapid, erratic variability.  Their sizes were measured by VLBI between Haystack in Massachusetts and Goldstone in California.  Both were most simply fit by two components of changing strength and changing separation (Whitney et al. 1971, Cohen et al. 1971, Shaffer et al. 1972), which looked like expansion.  If so, then the components were moving, it seemed, at twice or more times the speed of light.  Should this have been expected?  Yes, because it had been predicted (Rees 1966, 1967) and can best be described as a manifestation of special relativity, not a violation.

But not all were happy. Dent (1972, 1973) continued to monitor the variability (from Haystack) and concluded that he could fit what had happened with a succession of uncorrelated outbursts, happening at a rate of one or two per year, popping up and down at random places in a compact (but not Schwarzschild radius) core.  Chains of supernovae or stellar collisions came to mind, and the picture was called the ``Christmas tree model;'' for the style of (American) holiday lights that flash off and on.  Like many foolish things that were originally American, lights of this sort have spread and decorated the restaurant of the conference center where the 2012 Texas Symposium took place in Sa\~o Paulo, Brazil.

Just when the superluminal/Christmas tree issue was settled is a bit fuzzy, but Kellermann \& Moran (2001) suggest that it was images using multiple baselines collected and published by Pearson et al. (1981).  The implied expansion velocity was about $10c$.  Since then, many observers have plotted behavior of many superluminal jets where the components break loose from a central core, accelerate, decelerate, bend, and twist.  M87 is particularly impressive because it is fairly close (Giovanni et al. 2011, Asada et al. 2011).

Silence in the hall suggested that many participants did not agree with me that this discovery and confirmation of relativistic jet motion in AGNs was the single most important 3-d contribution of interferometry to the study of galaxies.  Perhaps narrowing the field to VLBI rather than all of radio interferometry would have increased agreement.  Changes with time of course constitute the 4th dimension in this story.

\section{Midcourse corrections and digressions}
Although we primarily associate Bernard Lovell with the 250-ft steerable paraboloid at Jodrell Bank, one of his earlier radio telescopes was an array of Yagis, working at a wavelength of 8.2 meters and designed to look for radar echoes from meteors. (Gunn 2005).  The first Japanese radio telescope was Tanaka's 1951 open mesh paraboloid, but by the 1970s there was an array of small, solid-surface dishes at Toyokawa (Kaifu 2013).  Grote Reber himself, though he constructed the first radio astronomy paraboloid, in due course built a sea cliff interferometer atop Mt. Haleakala, and, after his moves to Tasmania and to very low frequencies (that is, long wavelengths) of necessity switched to extended arrays of dipoles and such.  But the much later 23-meter University of Tasmania parabolic dish actually looks a good deal like the primordial Wheaton, IL installation (Kellermann 2005).

The Netherlands, having started with a W\H{u}rzburg 7.5 meter German radar dish just after the war, to look for predicted 21 cm radiation from neutral hydrogen, built the 25-m Dwingeloo paraboloid for that purpose, but then developed the Westerbork Synthesis Radio Telescope, an array of open-mesh paraboloids (Strom 2005).  They operate both to this day.  Australia also had a mix of technologies starting very early.

Asada et al. (2013) have suggested reconfiguring an ALMA test paraboloid and placing it in Greenland.  Operated in tandem with SMA, SMTO, LMT, IRAM, and ALMA itself this would provide 10 new arcs in the uv plane, extending to nearly 9000 km in the north-south (v) direction.  Coverage at the longest baselines will once again be sparse, as it was in the early days of VLBI.  And there is obviously a price to pay in having to observe sources very close to the horizon, and having large portions of the sky permanently invisible.

\section{High angular resolution radio instrumentation in the Decadal Surveys}
There have now been six of these surveys, generally named for the chairs of the main panels and titled, or subtitled, ``Astronomy for the 1980s'' and so forth,  There were early calls for more and larger paraboloids (Leo Goldberg said at one point that 500 feet was not too large for studying solar emission) and also for support of moderate sized paraboloids at universities.  Some of these things happened (See Trimble 2011 for details); some did not.  Table 1 shows the requests for interferometers and, more or less, what became of them.  The VLA clearly stands out, for prompt completion the first time around and the later requests for extensions, expansions, and incorporation into a VLBI network.  In its various incarnations, it has been, for most of its life, the most productive (of papers and citations) radio telescope in the world.  The VLBA has not, perhaps, been as successful as was expected.  Interferometry with one or more of the collecting areas in space was requested repeatedly, but the territory eventually handed over to our Japanese and Russian colleagues.  The Japanese HALCA lost one of its wavebands; the Russian RadioAstron, after many delays, was launched in July 2011.
\\
\\
\begin{table}[htdp]
\centering
\caption{Interferometry in the US Decadal Surveys}
\label{tab1}
\begin{tabular}{lll}\hline 
\\
{Report} & {Requests and Outcomes} 
\\ 
\\
{1964 Whitford} & {Pencil beam array, 100 x 85$^{\prime}$ dishes, to 3 cm (no)} \\ & {OVRO expanded to 6-8 dishes (eventually, as CARMA)}
\\
\\
{1972 Greenstein} & {VLA highest priority (construction began 1978)} \\ & {large mm array (ARO, ALMA)} \\ & {large cm array (VLA eventually pushed there)}
\\
\\
{1982 Field} & {VLB Array (1992)} \\ & {Space VLBI (to Japan, Russia)}
\\
\\
{1991 Bahcall} & {Millimeter array (ALMA, APEX, ASTE)} \\ & {Extended VLA (2012)} \\ & {Space radio interferometry}
\\
\\
{2001 McKee-Taylor} & {Expanded VLA (2012)} \\ & {SKA (MeerKAT, ASKAP, not US)} \\ & {BIMA merge with OVRO (CARMA 2007)} \\ & {Space radio interferometry} \\ & {LOFAR (NL dedicated 2010)} 
\\
\\
{2010 Blandford} & {No explicit requests on high priority list}
\\
\\
\hline
\end{tabular}
\vspace{1mm}
\end{table}

On the paraboloid side, the Whitford report asked for two 300$^{\prime}$ fully-steerable dishes operable down to 3 cm.  Whether the Greenbank 300$^{\prime}$  (commissioned in 1965) and the Goldstone 210$^{\prime}$  (mostly used for satellite tracking) met this goal is arguable.  They also requested a design study for the largest possible steerable paraboloid, which never really happened.  Bahcall asked for a 300$^{\prime}$  radio telescope in Brazil, after which filled dishes slipped to lower priority.

\section{Some (relatively) recent instruments and results}
One can easily collect a dozen or more acronyms representing recent, current, or planned radio interferometers of various sorts.  The VLA has become the J (for Janksy) VLA, BIMA and OVRO have merged to become CARMA.  And there are the SMA, PdB, IRAM, GMRT, ALMA, LOFAR, SKA, MeerKat, ACTA, WSRT, MERLIN, EVN, JVN, JCMT + CSO, and the VLBA  Many of these, including CARMA, PdB, SMA, and installations at Nobeyama and Stanford look very much like the VLA --- two-dimensional patterns of paraboloids (that would individually have seemed very large to Reber in 1932) on the ground, often on rail lines to permit changing of baselines more or less continuously, while the earth rotates providing lots more coverage of the u-v plane.  Some others --- LOFAR starting in the Netherlands, the SKA prototype ASKAP, the long wavelength array (LWA) near the VLA, and the MWA do not look like the VLA, and Kellermann \& Moran (2001) foresee a future in which digital processing can turn very large numbers of small receiver elements into interferometric phased arrays that can produce maps of large sky regions at many wavelengths all more or less at once.  The optical analogy is, I think, the IFU, much discussed at this meeting, but I wouldn't bet money on it.

From 1991 to 2006, I (and a few long-suffering friends) provided yearly reviews of the entire literature of astrophysics, mostly in PASP, and all published as Trimble or Trimble \& friend(s).  In the first nine years, the following interferometric, 3-d items were picked as highlights:
\\
\begin{itemize}
\item Bipolar outflows from galactic nuclei common and a signature of activity
\item Orientation of jets and unified models of AGNs
\item Many of the jets superluminal, not all (in sense of Sect. 6)
\item A (very) few counterrotating gas disks
\item A (very) few leading spiral arms
\item Galaxies with 1, 2, and 4 arms (the Milky Way still undecided)
\item Use of HALCA and RadioAstron for space VLBI
\item Gas structure in Milky Way core with two-sided jet and 3-armed spiral
\item Issue of correlation or anti-correlation of bright spots in jets between radio and optical still undecided (a registration problem, of some importance for understanding shock acceleration of relativistic electrons)
\item One shrinking superlumnial source
\item Resolution of lensed structures
\item Polarization of jets parallel vs perpendicular to jet length
\item Magellanic-Stream-like gas in other galaxies
\item Double bars
\item Maps of non-spherical accretion through cosmic web structures
\end{itemize}
In the present century, interferometric results have continued to pour forth, though my choice of ones to highlight may well seem perverse.  First was a WSRT HI image of NGC 4244, an edge-on disk galaxy that optically looks a good deal like NGC 891.  But NGC 891 has lots of HI out of its plane, including what seem to resemble the Milky Way's High Velocity Clouds.  NGC 4244 on the other hand has very blobby HI structure in a very thin plane and almost nothing at $z$ more than a couple hundred pc.  From a suitable point of view, almost a one-dimensional galaxy.

Then there is Arp 220 imaged with ALMA (Scoville et al. 2013).  The 349 GHz continuum image seems to have returned to us to a bipolar cow.  Ah, but it was really the spectrum they wanted, and, with the frequency selectivity of the array, they were able to pull out a line profile of HCN (4 - 3) that reveals H26 alpha on its low-frequency wing.

Ueda et al. (2013 in Kawabe et al. 2013, p. 61) have imaged 37 galaxies in CO that they describe a merger remnants, with CARMA, ALMA, SMA, and PdB.  No two are alike.  There are single central blobs, long worms, extended ellipses, a couple of bipolar cows, some apparent partial spirals, and a few total non-detections.  Someone else's IRAM image of 3C84 shows fairly clearly that the radio jets are pushing the CO outward into a couple of crescents.  And Espada (2013), who has mapped CO in the inner couple of kpc of Cen A has found a disk with spiral arms.  Cen A is, of course, a giant elliptical galaxy, which traditionally, therefore, should not have spiral arms.  Perhaps it belongs in with Ueda et al.'s merger remnants.

To bring these lists further up to date, please consult the rest of these proceedings and the descriptions of the sessions planned for the IAU General Assembly in Hawaii in August, 2015!
\\
\\
Acknowledgements\\
I am grateful to the organizers for the invitation to review the topic of history of interferometers and galactic structure and for partial support of my travel expenses.  The rest, of course, comes from The Government, in the form of a Schedule A tax deduction.  Ms. Alison Lara turned my usual scruffy typescript into a submittable file in her usual expert fashion.

\end{document}